\documentclass[aps,pre,twocolumn,floats,amsmath,amssymb,showpacs,floatfix]{revtex4-1}
\usepackage{dcolumn}
\usepackage{subfigure}
\usepackage{psfrag,graphicx, epsfig,color}

\begin{document}
\title{Triangular Constellations in Fractal Measures}
\author{Michael Wilkinson and John Grant}
\affiliation{Department of Mathematics and Statistics,
The Open University, Walton Hall, Milton Keynes, MK7 6AA, England}

\begin{abstract}
The local structure of a fractal set is described
by its dimension $D$, which is the exponent
of a power-law relating the mass ${\cal N}$ in a ball
to its radius $\epsilon$: ${\cal N}\sim \epsilon^D$.
It is desirable to characterise the {\em shapes} of constellations
of points sampling a fractal measure, as well as their masses.
The simplest example is the distribution of shapes of triangles formed by triplets
of points, which we investigate for
fractals generated by chaotic dynamical systems. The most
significant parameter describing the triangle shape is the
ratio $z$ of its area to the radius of gyration squared. We show that
the probability density of $z$ has a phase transition: $P(z)$ is
independent of $\epsilon$ and approximately uniform
below a critical flow compressibility $\beta_{\rm c}$, but for $\beta>\beta_{\rm c}$
it is described by two power laws: $P(z)\sim z^{\alpha_1}$ when $1\gg z\gg z_{\rm c}(\epsilon)$, and  $P(z)\sim z^{\alpha_2}$ when $z\ll z_{\rm c}(\epsilon)$.
\end{abstract}

\maketitle

Fractal sets and measures play a pivotal role in many areas of physics
\cite{Gou96}.
Fractals are characterised by exploring their local structure.
Consider, for example, a set of points obtained by sampling a fractal measure
(these could be points representing trajectories
in a phase space with a chaotic attractor).
One commonly used approach is to pick one of the points at
random, and then investigate the number of other points inside
a sphere of radius $\epsilon$ centred on that test point. If the
expectation value of the number of points ${\cal N}$ is a power-law
in $\epsilon$,
\begin{equation}
\label{eq: 1}
\langle {\cal N}(\epsilon)\rangle \sim \epsilon^{D_2}
\end{equation}
then $D_2$ is the correlation dimension of the set \cite{Gra+84} (throughout this paper
$\langle X\rangle$ denotes the ensemble average of $X$). It is a characteristic
feature of fractals that their local structure is characterised by power-laws,
such as (\ref{eq: 1}). More general definitions of dimension, involving different
moments of the mass, are discussed in \cite{Hal+86}.

Fractals which have nearly identical values of the dimension can have a very
different appearance. It is desirable to develop means to characterise
the shape of the internal structure of fractal distributions, because
differences in the local structure of fractal sets may have important
implications for properties such as light scattering or network
connectivity.

Here we address the simplest question about the
internal shape-structure of a fractal set. Consider two randomly
chosen particles in a ball of radius $\epsilon$ surrounding a
reference point. Together with the test
point, these define a triangle. The local structure can be described
in greater detail by specifying the statistics of the shapes of
these triangles. The shape of a triangle is described by a point in a
two-parameter space (we could choose two of the angles, but
a better choice is described later).

Many point-set fractals arise as a result of dynamical processes: examples
are strange attractors \cite{Ott02}, distributions of
particles in turbulent flow \cite{Som+93}, and possibly
also the distributions of matter resulting from gravitational collapse \cite{Pie87}.
In this paper we analyze the distribution of triangle
shapes for a generic model of chaotic dynamics.
We show that the distribution of
triangle shapes is also associated with power-laws.
It might be expected that the stretching action of the dynamics will
exaggerate the prevalence of thin, acute-angled triangles. This expectation
is only partially correct.

The statistics of the shapes of triangles drawn from a random
scatter of points was addressed by Kendall \cite{Ken77}.
He showed that there is a natural parametrisation of
the shape of a triangle in terms of a point on the surface of a sphere,
with equilateral triangles of opposite orientation at the poles,
and with degenerate triangles consisting of co-linear points lying
on the equator. Kendall observed that the image of a Brownian motion
of the three corners is a Brownian motion on the surface of the sphere
(a transparent demonstration of this result is given in \cite{Pum+13}).

Our discussion of the triangle shapes will emphasise the coordinate
$z=\cos\Theta$ where $\Theta$ is the polar angle on Kendall's sphere. This
is related to the area ${\cal A}=\delta\mbox{\boldmath$r$}_1\wedge\delta\mbox{\boldmath$r$}$
and the radius of gyration ${\cal R}$:
\begin{equation}
\label{eq: 2}
z=\frac{\cal A}{\sqrt{3}{\cal R}^2}
\ ,\ \ \
{\cal R}^2=\frac{1}{6}\left[(\delta \mbox{\boldmath$r$}_1)^2+(\delta \mbox{\boldmath$r$}_2)^2
+(\delta \mbox{\boldmath$r$}_1-\delta \mbox{\boldmath$r$}_2)^2\right]
\end{equation}
where the $\delta \mbox{\boldmath$r$}_i$ are displacements of two
points relative to the third, reference, point.
For a random scatter of points, Kendall showed that $z$ has a uniform
distribution on $[-1,1]$: $P(z)=\frac{1}{2}$.
Note that thin, acute triangles correspond to small values of $z$.
We concentrate upon the distribution $P(z)$ in the limit as $z\to 0$.

\begin{figure*}[t]
\includegraphics[width=1.0\textwidth]{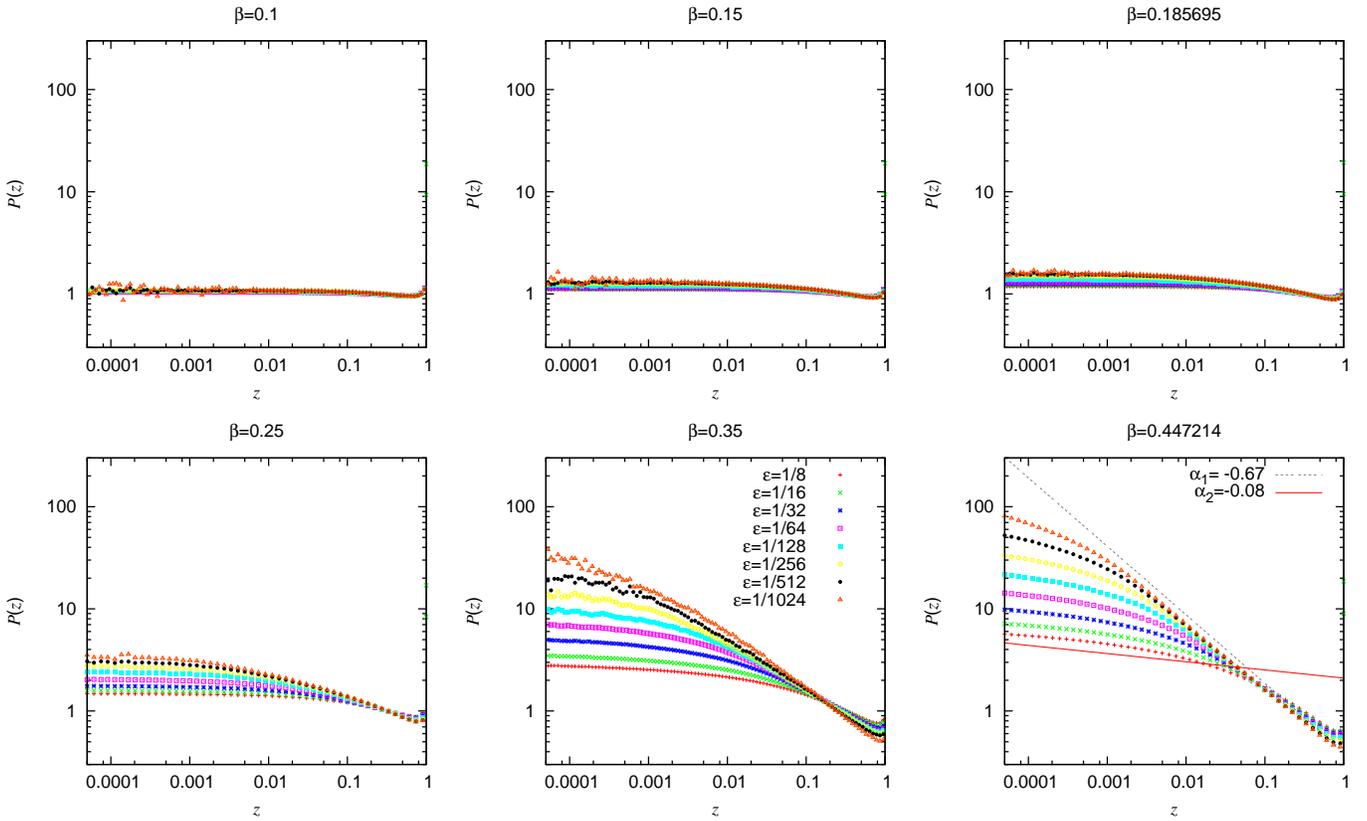}
\caption{\label{fig: 1}(Colour online). Probability density $P(z)$ for various
compressibilities $\beta$: $\beta_{\rm c}=1/\sqrt{29}=0.185\ldots$ is our
estimate of the critical compressibility, and $D_2=1$ when $\beta=1/\sqrt{5}=0.447\ldots$.
Straight lines indicate estimates for $\alpha_1$ and $\alpha_2$ when $\beta>\beta_{\rm c}$.
Note that $P(z)$ is normalisable
($\alpha_1>-1$) even at $\beta=1/\sqrt{5}$, where $D_2=1$.}
\end{figure*}

As a concrete example of a dynamical process which generates a fractal measure we
consider particles advected in a random flow in two dimensions
\cite{Som+93}. The analysis is readily adapted to other
dynamical systems. The equation of motion is
$\dot{\mbox{\boldmath$r$}}=\mbox{\boldmath$u$}(\mbox{\boldmath$r$},t)$
where $\mbox{\boldmath$u$}(\mbox{\boldmath$r$},t)$ is a random velocity field.
We consider only particles which are
sufficiently close that their separation has a linear equation of motion
defined by a $2\times 2$ matrix ${\bf A}$ with elements $A_{ij}$:
\begin{equation}
\label{eq: 3}
\delta \dot{\mbox{\boldmath$r$}}={\bf A}(t)\delta \mbox{\boldmath$r$}
\ , \ \ \
A_{ij}(t)=\frac{\partial u_i}{\partial r_j}(\mbox{\boldmath$r$}(t),t)
\ .
\end{equation}
Figure \ref{fig: 1} shows the numerically determined distribution
of $z$ for small triangular constellations formed by
triplets of randomly chosen points inside a disc of radius
$\epsilon\ll \xi$, where $\xi=0.25$ is the correlation length of the flow.
The plots show the probability distribution $P(z)$ for particles
advected in six different random flows, with differing degrees of
compressibility, described by a parameter $\beta$. In each case
the distributions for eight different values of $\epsilon$ are
shown on double-logarithmic scales. The model flow is described in detail
below. For small compressibility $\beta$, the distribution is
approximately independent of the value of $\epsilon$ and uniform
(apart from a cusp at $z=1$ which arises because our sampling criterion
is slightly different from Kendall's). When $\beta>\beta_{\rm c}$,
$P(z)$ becomes dependent upon $\epsilon$ and is asymptotic
to two power laws in the limit as $\epsilon \to 0$:
$P(z)\sim z^{\alpha_1}$ when $z$ is small, but exceeds
a value $z_{\rm c}(\epsilon)$ which decreases as
$\epsilon \to 0$, and $P(z)\sim z^{\alpha_2}$ for $z\ll z_{\rm c}$.
In the remainder of this paper we explain why $P(z)$ has power-law
behaviour, why there is a critical compressibility $\beta_{\rm c}$,
and why the distribution has two exponents for $\beta>\beta_{\rm c}$.

The size and shape of a triplet of points $(\mbox{\boldmath$r$}_0,\mbox{\boldmath$r$}_1,\mbox{\boldmath$r$}_2)$
may be described by three parameters: $R_1$, $R_2$, and $\delta \theta$, which are defined by parametrising the separations $\delta \mbox{\boldmath$r$}_i=\mbox{\boldmath$r$}_i-\mbox{\boldmath$r$}_0$ as follows:
\begin{equation}
\label{eq: 4}
\delta \mbox{\boldmath$r$}_1=R_1{\bf n}_1
\ ,\ \ \
\delta \mbox{\boldmath$r$}_2=R_2({\bf n}_1+\delta \theta {\bf n}_2)
\end{equation}
where ${\bf n}_1$, ${\bf n}_2$ are two orthogonal, time dependent unit
vectors and the particles are labelled such that $R_1\ge R_2$.
The equations of motion for $R_1$, $R_2$ and $\delta \theta$ are
obtained by substituting (\ref{eq: 4}) into (\ref{eq: 3}) and
projecting the equations of motion for $\delta \mbox{\boldmath$r$}_i$ onto
the ${\bf n}_j$. Using the notation
\begin{equation}
\label{eq: 5}
F_{ij}={\bf n}_i\cdot {\bf A}\,{\bf n}_j
\end{equation}
we obtain $\dot {\bf n}_1\cdot {\bf n}_2=F_{21}(t)$,
\begin{equation}
\label{eq: 6}
\frac{\dot R_1}{R_1}=F_{11}(t)
\ ,\ \ \
\frac{\delta \dot \theta}{\delta \theta}=F_{22}(t)-F_{11}(t)
\end{equation}
and a similar equation for $\dot R_2/R_2$.

The important point about (\ref{eq: 6}) is that the logarithmic derivative
is expressed in terms of the randomly fluctuating quantities $F_{ij}(t)$,
so that it is advantageous to transform to logarithmic
variables. The variables $(R_1,R_2,\delta\theta)$ can be replaced by
\begin{equation}
\label{eq: 7}
X_1=-{\rm ln}\, \frac{R_1}{\xi}
\ ,\ \ \
X_2=-{\rm ln}\, \delta \theta
\ ,\ \ \
X_3={\rm ln}\, \left(\frac{R_1}{R_2}\right)
\end{equation}
where $\xi$ is the correlation length of the velocity field.
When $X_1$ and $X_2$ satisfy $X_1\gg 0$ and $X_2\gg 0$,
(that is we are dealing with a small, acute-angled triangles),
the dynamics is trivial: $X_3$ is frozen and $X_1$, $X_2$ obey
stochastic equations of motion $\dot X_i=\eta_i(t)$,
where the $\eta_i(t)$ are random functions of time, with statistics
independent of position in $(X_1,X_2,X_3)$ space. Because
$X_3$ is frozen, we have $\eta_3(t)=0$. The probability density $P(X_1,X_2,X_3)$
obeys a steady-state advection-diffusion equation,
$v_i\partial_i P+ {\cal D}_{ij}\partial_i\partial_j P=0$
(with $\partial_i=\partial/\partial X_i$, and summation over
repeated indices). The drift velocities, correlation
functions and diffusion coefficients are:
\begin{eqnarray}
\label{eq: 8}
v_i=\langle \dot X_i(t)\rangle
\ \ \ &&\ \ \
C_{ij}(t)=\langle [\dot X_i(t)-v_i][\dot X_j(0)-v_j]\rangle
\nonumber \\
{\cal D}_{ij}&=&\frac{1}{2}\int_{-\infty}^\infty {\rm d}t\ C_{ij}(t)
\ .
\end{eqnarray}

At this point we can already see why $z$ may have
a power-law distribution.
Note that, because $z\sim \delta \theta$, if $\delta \theta$ has a
power-law distribution, then $z$ also has a power-law distribution
with the same exponent. Because the equation of
motion for $(X_1,X_2,X_3)$ is translationally invariant
in the sector $X_1,X_2\gg 0$, any function invariant under translation (up to a change of normalisation)
gives a steady-state solution for $P(X_1,X_2,X_3)$.
Because the exponential function is translationally invariant,
solutions exist in the form $P(X_1,X_2,X_3)=\exp(\gamma_1 X_1+\gamma_2 X_2)P_3(X_3)$ (where
normalisation requires that the constants $\gamma_i$ are negative).
The corresponding distributions of $R_1$ and $\delta \theta$ have
probability densities proportional to $R_1^{\gamma_1-1}$ and
$\delta\theta^{\gamma_2-1}$ respectively, so that $P(z)\sim z^\alpha$ with
$\alpha =\gamma_2-1$. Also, we identify $R_1$ with $\epsilon$ and note that
$\epsilon^{\gamma_1}$ is the probability density for both points being within
a ball of radius $\epsilon$ surrounding the reference point, implying that $\gamma_1=2D_3$,
where $D_3$ is the third Renyi dimension \cite{Hal+86}.

In order to identify which power-law solutions obtain we must consider
the boundary conditions on the lines $X_1=0$ and $X_2=0$.
The boundary $X_1=0$ is a distributed
source, corresponding to random triplets of points with a separation
approximately equal to the correlation length $\xi$. Some of these are
\lq squeezed' by the linearised flow so that they enter the region $X_1>0$.
Phase points $\mbox{\boldmath$X$}$ representing these triplets are created on the line $X_1=0$ at a rate
$J(X_2)$ which corresponds to $\delta \theta$ having a uniform
probability density in the limit as $\delta \theta\to 0$ (as required by Kendall's result \cite{Ken77}),
implying that the source density on the boundary $X_1=0$ is
\begin{equation}
\label{eq: 9}
J(X_2)=\left\{
\begin{array}{cc}
J_0\exp(-X_2) & X_2<0 \cr
0            & X_2\ge 0
\end{array}
\right .
\end{equation}
(where $J_0$ is a constant, which determines the normalisation
of the joint probability density, $P(X_1,X_2,X_3)$).
The boundary at $X_2=0$ is more complicated: it is non-absorbing,
and the approximations that $X_3$ decouples and $(X_1,X_2)$ obey
a simple diffusion-advection equation fail close to
$X_2=0$.

It is not possible to determine the steady-state
solution of the advection-diffusion equation exactly.
In the following we use an approximate propagator
to make a quantitative theory for the critical compressibility
$\beta_{\rm c}$, and give a qualitative explanation of the reason
why there are two exponents, $\alpha_1$ and $\alpha_2$.
Because the probability density is expected to vary over a wide range of values,
it is conveniently expressed using an exponential form
\begin{equation}
\label{eq: 10}
P(\mbox{\boldmath$X$})\sim \exp\left[-\Phi(\mbox{\boldmath$X$})\right]
\ .
\end{equation}
Consider first the form of the exponent $\Phi(\mbox{\boldmath$X$})$ for the advection-diffusion
equation with drift velocity $\mbox{\boldmath$v$}$ and diffusion tensor ${\bf D}$
when there is a point source at the origin. In this case the exponent in (\ref{eq: 10})
will be denoted by $\Phi_0(\mbox{\boldmath$X$})$. If the source is localised at
$\mbox{\boldmath$X$}={\bf 0}$ and at $t=0$, the solution of the diffusion process in
$d$ dimensions is a gaussian centred at $\mbox{\boldmath$X$}=\mbox{\boldmath$v$}t$.
The steady-state probability density from a constant intensity source at
$\mbox{\boldmath$X$}={\bf 0}$ can be obtained by integration of this gaussian
propagator over $t$:
\begin{eqnarray}
\label{eq: 11}
P(\mbox{\boldmath$X$})&=&K\int_0^\infty {\rm d}t\ \left[4\pi {\rm det}({\bf D})t\right]^{-d/2}
\exp\left[-S(\mbox{\boldmath$X$},t)\right]
\nonumber \\
S(\mbox{\boldmath$X$},t)&=&
\frac{1}{4t}(\mbox{\boldmath$X$}-\mbox{\boldmath$v$}t)
\cdot{\bf D}^{-1}(\mbox{\boldmath$X$}-\mbox{\boldmath$v$}t)
\ .
\end{eqnarray}
Here $K$ is a normalisation constant. To estimate $\Phi_0(\mbox{\boldmath$X$})$,
we determine the time $t^\ast$ where the
propagator is maximal, and set
$\Phi_0(\mbox{\boldmath$X$})=S(\mbox{\boldmath$X$},t^\ast)$,
where $\partial S/\partial t(\mbox{\boldmath$X$},t^\ast)=0$.
The equation for $t^\ast$ is
$\mbox{\boldmath$X$}\cdot{\bf D}^{-1}\mbox{\boldmath$X$}
-t^{\ast 2}
\mbox{\boldmath$v$}\cdot{\bf D}^{-1}\mbox{\boldmath$v$}=0$.
Hence
\begin{equation}
\label{eq: 12}
\Phi_0(\mbox{\boldmath$X$})=\frac{1}{2}\left[\sqrt{\mbox{\boldmath$X$}\
\cdot{\bf D}^{-1}\mbox{\boldmath$X$}}\sqrt{\mbox{\boldmath$v$}\
\cdot{\bf D}^{-1}\mbox{\boldmath$v$}}
-\mbox{\boldmath$X$}\cdot{\bf D}^{-1}\mbox{\boldmath$v$}
\right]
\ .
\end{equation}
Note that $\Phi_0(X_1,X_2)$ is the height of a tilted conical surface, which
touches the $(X_1,X_2)$ plane along the ray $\mbox{\boldmath$X$}=\lambda \mbox{\boldmath$v$}$
with parameter $\lambda$ (\lq downwind' of the source), but increases
linearly along any other ray starting from the source point.
Correspondingly, there is an asymptotically exponential reduction of $P(X_1,X_2)$ along
any ray from the source.

Next we estimate $P(\mbox{\boldmath$X$})=\exp[-\Phi(\mbox{\boldmath$X$})]$ taking
account of the condition that there is a distributed source
on the line $X_2>0$, with intensity $J(X_2)=\exp(-X_2)$.
The probability is obtained by integrating the point-source
solution $\exp[-\Phi_0(\mbox{\boldmath$X$})]$ over the initial point,
$(X_1,X_2)=(0,X_0)$, with $X_0>0$:
\begin{equation}
\label{eq: 13}
P(X_1,X_2)=\int_0^\infty {\rm d}X_0\ \exp(-X_0) \exp[-\Phi_0(X_1,X_2-X_0)]
\ .
\end{equation}
We may assume that the integral is dominated by contributions
from the vicinity of a stationary point $X^\ast$. The exponent in (\ref{eq: 11})
is then determined by
\begin{eqnarray}
\label{eq: 14}
\Phi(X_1,X_2)&=&\Phi_0(X_1,X_2-X^\ast)+X^\ast
\nonumber \\
0&=&1-\frac{\partial \Phi_0}{\partial X_2}(X_1,X_2-X^\ast)
\ .
\end{eqnarray}
Because of the conic structure of the function $\Phi_0(X_1,X_2)$,
the gradient of this function is constant along any ray from the source
point $(X_1,X_2)=(0,X^\ast)$. For any given value of the compressibility
parameter $\beta$ the stationary point condition is satisfied on a ray which
leaves the source point with a slope $s(\beta)$ (which is determined for a specific model below).
The stationary point is therefore
\begin{equation}
\label{eq: 15}
X^\ast=X_2-s(\beta)X_1
\ .
\end{equation}
Hence $\Phi(X_1,X_2)=\Phi_0(X_1,s(\beta)X_1)+X_2-s(\beta)X_1$.

This analysis is only correct if there is a valid stationary point, that
is when (\ref{eq: 15}) predicts $X^\ast>0$. If $s(\beta)<0$, there is always
a positive solution to equation (\ref{eq: 15}), and $\Phi(X_1,X_2)\sim X_2$,
so that $P(z)\sim z^0$. If $s(\beta)>0$, however, stationary points (\ref{eq: 15})
do not exist when $(X_1,X_2)$ lies below the line of slope $s(\beta)$.
In this case the integral (\ref{eq: 13}) is dominated by the contribution
from $X_0=0$, and we have $\Phi(X_1,X_2)\sim \Phi_0(X_1,X_2)$.
In the case where $s(\beta)>0$, there are two exponents ($\alpha_2$ and $\alpha_1$
respectively) which characterise
$P(z)$, depending upon whether $(X_1,X_2)=({\rm ln}(\xi/\epsilon),-{\rm ln}(\delta \theta))$
lies above or below the line of slope $s(\beta)$. These exponents are, respectively,
$\alpha_1=\partial_{X_2} \Phi (X_1,0)-1$,
$\alpha_2=\lim_{X_2\to \infty}\partial_{X_2} \Phi(X_1,X_2)-1$.

The phase transition illustrated
in figure \ref{fig: 1} is determined by the condition $s(\beta_{\rm c})=0$.
Consider the source of small, non-acute triangles
(represented by $X_1\gg 1$, $X_2\approx 0$). When $\beta<\beta_{\rm c}$,
there is a solution of (\ref{eq: 14}) and
these triangles are predominantly formed by squeezing of
acute triangles from $(0,X^\ast)$ along their axis. When $\beta>\beta_{\rm c}$,
they are formed by
approximately isotropic squeezing of non-acute triangles.

We now show how these predictions are illustrated by the numerical
results plotted in figure \ref{fig: 1}. In our numerical
investigations we have used the map
\begin{equation}
\label{eq: 16}
\mbox{\boldmath{$x$}}_{n+1}=\mbox{\boldmath{$x$}}_n+\mbox{\boldmath{$u$}}_n(\mbox{\boldmath{$x$}}_n)
\sqrt{\delta t}
\end{equation}
with the velocity field $\mbox{\boldmath$u$}$ constructed from a stream function $\psi$ and a scalar potential $\phi$:
\begin{equation}
\label{eq: 17}
\mbox{\boldmath$u$}_n=\left(\partial_y \psi_n+\beta \partial_x \phi_n,-\partial_x\psi_n+\beta\partial_y\phi_n\right)
\ .
\end{equation}
This is equivalent to using a velocity field which is
delta-correlated in time \cite{Fal+01}. The timestep $\delta t$ is
assumed to be sufficiently small to allow the use
of diffusive approximations.
The random fields $\phi$ and $\psi$ have the same translationally invariant and
isotropic statistics. They are independent of each other and chosen independently
at each timestep, and were normalised so that
$\langle (\partial_{xy}\psi)^2\rangle=1$. For this flow the elements
of the diffusion tensor ${\bf D}$ are ${\cal D}_{11}=\frac{1}{2}(1+3\beta^2)$,
${\cal D}_{12}={\cal D}_{21}=-(1+\beta^2)$, ${\cal D}_{22}=2(1+\beta^2)$
and the components of the drift velocity are $v_1=1-\beta^2$, $v_2=-2(1+\beta^2)$.
The correlation dimension is $D_2=-v_1/{\cal D}_{11}=2(1-\beta^2)/(1+3\beta^2)$,
\cite{Bec+04,Wil+12} (so that $D_2=1$ for $\beta=1/\sqrt{5}$). Writing
$K=(1+3\beta^2)/(2\beta\sqrt{2(1+\beta^2)})$, $\Lambda=4(1+\beta^2)/(1+3\beta^2)$,
the exponent (\ref{eq: 12}) for the point-source solution is
\begin{equation}
\label{eq: 18}
\Phi_0(\mbox{\boldmath$x$})=K
\sqrt{X_2^2+\Lambda(X_1^2+X_1X_2)}
-X_1-X_2
\ .
\end{equation}
Substituting this into (\ref{eq: 14}), we find that the stationary point
satisfies (\ref{eq: 15}) with slope
\begin{equation}
\label{eq: 19}
s(\beta)=\frac{2(1+\beta^2)}{1+3\beta^2}\left[\frac{8\beta^2}{\sqrt{1-26\beta^2-23\beta^4}}-1\right]
\end{equation}
so that $s(\beta)=0$ at $\beta=1/\sqrt{29}=0.185\ldots $, where $D_2=7/4$.
This implies that $P(z)\sim z^0$, independent
of $\beta$, for $\beta \le \beta_{\rm c}=1/\sqrt{29}$. This prediction for the
critical compressibility is in good agreement with our numerical results,
illustrated in figure \ref{fig: 1}. When $\beta>1/\sqrt{29}$, the there are
two different exponents for $P(z)$ because the dominant
contribution to the propagator depends upon the position in the $(X_1,X_2)$ plane.
We remark that a quantitative treatment of the exponents $\alpha_1$, $\alpha_2$ when
$\beta >\beta_{\rm c}$ requires a more sophisticated model for the propagator, taking
account of the different boundary condition at $X_2=0$.

In conclusion, we investigated the distribution of shapes of triangular constellations
in fractal sets arising from compressible chaotic flow. We find that the distribution
of the acute angle is approximately independent of compressibility $\beta$ up to a critical point
$\beta_{\rm c}$, which we were able to determine analytically. For $\beta>\beta_{\rm c}$
the exponent $\alpha$ in $P(z)\sim z^\alpha$ takes two different values,
depending upon how small $z$ is.
We thank Alain Pumir for helpful discussions.

\end{document}